\documentclass[aps,pra]{revtex4}

\usepackage{epsfig,pstricks,dina4}

\usepackage{graphicx}
\usepackage{amsmath,amssymb}

\begin{document}

\def\DJo{$\;$\kern-.4em \hbox{D\kern-.8em\raise.15ex\hbox{--}\kern.35em okovi\'c}}
\def\CC{{\rm\kern.24em \vrule width.04em height1.46ex depth-.07ex
\kern-.30em C}}
\newcommand{\beq}{\begin{equation}}
\newcommand{\beqa}{\begin{eqnarray}}
\newcommand{\eeq}{\end{equation}}
\newcommand{\eeqa}{\end{eqnarray}}
\newcommand{\bra}[1]{\left\langle #1 \right |}
\newcommand{\ket}[1]{\left | #1 \right\rangle}
\def\trace{{\rm tr}\;}
\def\e{{\rm e}}
\newcommand{\eps}{\varepsilon}
\newcommand{\bigfrac}[2]{\mbox {${\displaystyle \frac{ #1 }{ #2 }}$}}
\newcommand{\Matrix}[2]{\left( \begin{array}{#1} #2 \end{array}
  \right)}
\newcommand{\diag}{{\rm diag}\;}

\renewcommand{\Im}{{\rm Im\;}}
\renewcommand{\Re}{{\rm Re\;}}
\def\RR{{\rm
         \vrule width.04em height1.58ex depth-.0ex
         \kern-.04em R}}
\def\id{{\rm 1\kern-.22em l}}

\title{Classification of qubit entanglement: $SL(2,\CC)$ versus $SU(2)$ invariance}
\author{Andreas Osterloh}
\affiliation{Institut f\"ur Theoretische Physik, Leibniz Universit\"at Hannover, D-30167 Hannover, Germany}

\begin{abstract}
The role of $SU(2)$ invariants for the classification 
of multiparty entanglement is discussed and exemplified for the 
Kempe invariant $I_5$ of pure three-qubit states. 
It is found to being an independent invariant
only in presence of both $W$-type entanglement and threetangle.
In this case, constant $I_5$ admits for a wide range of both threetangle
and concurrences. Furthermore, the present analysis indicates that an
$SL^{\otimes 3}$ orbit of states with equal tangles but 
continuously varying $I_5$ must exist.
This means that $I_5$ provides no information 
on the entanglement in the system in addition to that contained in 
the tangles (concurrences and threetangle) themselves.
Together with the numerical evidence that $I_5$ is an entanglement monotone
this implies that $SU(2)$ invariance or the monotone property
are too weak requirements for the characterization and quantification of 
entanglement for systems of three qubits, and that $SL(2,\CC)$ invariance 
is required. This conclusion can be extended to general multipartite systems 
(including higher local dimension) because the entanglement classes 
of three-qubit systems appear as subclasses.
\end{abstract}

\maketitle

\section{Introduction}
The understanding of entanglement is a central issue at the heart 
of quantum information theory.
First insight has been obtained for
the archetype of entanglement, namely the bipartite case, and in 
particular for binary observables of the constituents
(e.g. photon polarization, two-level systems) hence termed qubits.
Many criteria have been conceived to distinguish disentangled from
entangled pure states as the Schmidt rank and the 
von Neumann entropy followed by varieties of measures for the 
mixedness of the local 
reduced density matrix\cite{NielsenK01,Zyczkowski02}. 
Necessary criteria for any entanglement measure to
be fulfilled have been elaborated and have lead to the notion of an 
{\em entanglement monotone}~\cite{MONOTONES}. Crucial
requirements are local $SU(2)$ invariance, convexity on the space
of density matrices and monotonic diminishment under local SLOCC
(Stochastic Local Operations and Classical Communication)~\cite{SLOCC}.
For pure two-qubit states an explicit extension to mixed states -- 
i.e. the {\em convex roof}\/\cite{Uhlmann00} has been found for the 
concurrence and derived measures~\cite{Hill97,Wootters98,Uhlmann00}.
Indeed, from the convex roof of the concurrence also the convex roof 
of any other measure of pairwise qubit entanglement can be obtained 
via its relation to the concurrence for pure states.

The success and the impressive achievements for the bipartite case 
asked for extensions to the multipartite setting and also to 
higher local dimensions. 
However, complications arise from the appearance of
various entanglement classes that are inequivalent under 
SLOCC. It has been understood that
SLOCC transformations lead to local transformations in the group $SL(2,\CC)$ rather than
just $SU(2)$.
Based on this insight one can proof that only one 
SLOCC class of global entanglement exists for three qubits\cite{Duer00}. 
Its representative is the GHZ-state and has genuine three-party entanglement,
as measured by the threetangle~\cite{Coffman00}. 
A representative of the zero SLOCC class is the $W$ state, which 
exclusively contains pairwise entanglement, as measured by the concurrence. 
However, a precise e.g. operational meaning of this classification 
is still missing:
neither a clear analogue to the entanglement of formation exists,
nor is it known, whether a finite set of entangled states existed from which
all pure states could be generated\cite{NoMREGS}.

Much work has been done in order to classify
multipartite entanglement for more than three qubits
~\cite{VerstraeteDMV02,Luque02,VerstraeteDM03,Miyake03,OS04,OS05,Luque05,Dokovic,LTT,OD08}, 
but it is still not clear what conditions are to be imposed on 
such a classification.
The minimal requirement is certainly local $SU(2)$ invariance, 
but an extension to the local $SL(2,\CC)$ group shows many appealing 
advantages. Although it is known in principle
how to generate invariants to a certain group~\cite{OLVER}, 
the {\em real} problem consists in the distillation of those invariants
relevant for entanglement. 
As to give an example, for four qubits
nine different SLOCC classes have been identified\cite{VerstraeteDMV02}
as opposed to $19$ primary and $1.449.936$ secondary $SU(2)$ invariants 
obtained from the Hilbert series\cite{LTT,BriandLT03}. 
On the other hand, only four inequivalent polynomial 
$SL(2,\CC)^{\otimes 3}$ invariants do exist, three of which 
vanish on all product states, and three inequivalent
maximally entangled states have been singled out, 
which are all grouped in one of rhe SLOCC-class proposed in 
Ref.~\cite{VerstraeteDMV02}.

In order to approach an answer to this problem, the three qubit case provides 
an ideal playground since a complete set of $SU(2)$ invariants is known 
to contain only $6$ independent elements. 
Only four of them are partially $SL(2,\CC)$ invariant,
and simple normal forms depending on $6$ real parameters are available.
They have been presented independently in Refs.~\cite{Acin00,Carteret00} 
as multipartite extensions to the Schmidt decomposition.
Both have been analyzed with respect to a complete characterization 
of the local unitary orbits~\cite{Acin01,Gingrich02}.
For this purpose, a canonical form of three qubit states 
from~\cite{Acin00} has been expressed in terms of polynomial $SU(2)$ invariants. 
Here, focus is given on the connection between the same canonical form and
the exhaustive classification of three qubit entanglement in \cite{Duer00}.
To this end, we present an expression of the Ac\'{\i}n normal form
in terms of four (partial) $SL(2,\CC)$ invariants: the
threetangle $\tau_3$ and the concurrences $C_{1,2}$, $C_{1,3}$, 
and $C_{2,3}$. This expression discriminates the merely $SU(2)$ invariants 
from $SL(2,\CC)$ invariants based on entanglement related questions. 
We argue that $SU(2)$ invariance alone is not sufficient in order to
quantify and classify the entanglement pattern of a state,
and that $SL(2,\CC)$ invariance must be present somewhere for this purpose. 
First weak evidence for this to hold is
(i) there is one $SU(2)$ invariant that trivially has nothing to do
with the entanglement of the state, namely its modulus.
So, for merely $SU(2)$ invariant quantities, in order to be
related with entanglement, at least eventual non-local symmetries must be excluded. 
(ii) The only further invariant for two qubits is the concurrence,
which is $SL(2,\CC)$ invariant.
This coincides with one existing entanglement class in this case. 

For three qubits there is one further merely $SU(2)$ invariant that is functionally 
independent from the four tangles: the Kempe invariant $I_5$\cite{Sudbery01,Kempe99}
\beq
I_5=3\trace(\rho_i\otimes\rho_j)\rho_{ij} - \trace\rho_i^3-\trace\rho_j^3\; .
\eeq
$I_5$ distinguishes locally indistinguishable states\cite{Kempe99}
and is related to the relative entropy of $\rho_{ij}$ and 
$\rho_i\otimes\rho_j$ of the two-qubit state\cite{Sudbery01}. 
It is permutation invariant but not with respect to $SL(2,\CC)$ on any qubit. 
We analyze this quantity on pure three qubit states in order to understand
the insight it gives into the entanglement of the state.
%%%%%%%%%%%%%%%%%%%%%%%%%%%%%
\vspace*{5mm}
\begin{figure}[ht]
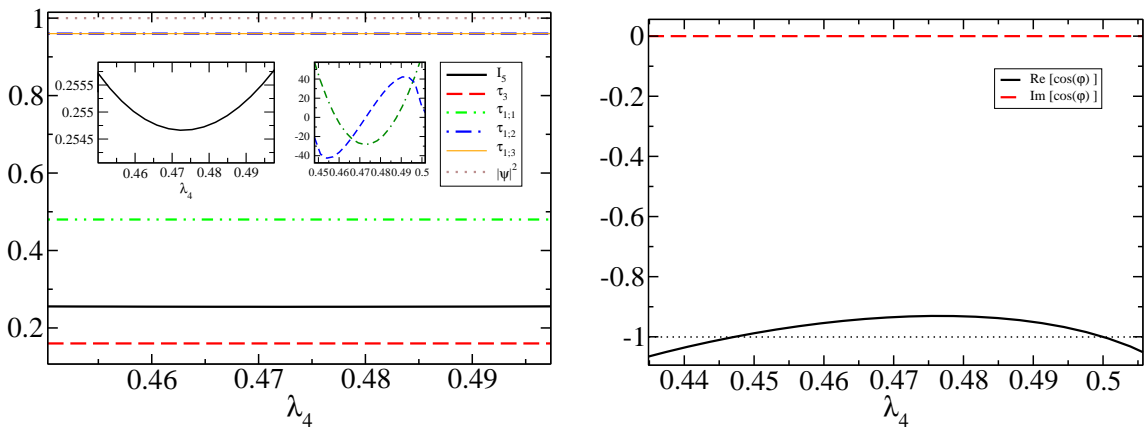

\begin{center}
\includegraphics[angle=0,width=0.48\textwidth]{I5.1111.pi.eps}
\hfill
\includegraphics[angle=0,width=0.48\textwidth]{Cosphi.1111.pi.eps}
\caption{{\em Left}: 
$I_5$ (black solid line) together with $\tau_3$ and the local 
tangles $\tau_{1;i}$ for the states in Ac\'{\i}n normal form with
the entanglement pattern of the state (\ref{example}) and $\alpha=\pi$.
The left inset shows that $I_5$ indeed varies, whereas the right inset shows the
behavior of the Grassl invariant~\cite{Acin01} $I_G$ (rescaled by $10^3$): 
the green dash-dash-dotted line is the real part of $I_G$; the blue dashed line its 
imaginary part.
{\em Right}: 
The interval of validity is given by the 
expression for $\cos\varphi$.
\label{Iso-pi}
}
\end{center}
\end{figure}
%%%%%%%%%%%%%%%%%%%%%%%%%%%%%
The paper is organized as follows. The next section discusses 
the functional independence of $I_5$ within a complete set of $SU(2)$
invariants. In Section~\ref{main} we obtain a one-parameter family of 
states in canonical form with all tangles (concurrences and threetangle) 
fixed and analyze this one-parameter family
showing the behavior of $I_5$ in the interval consistent with
the fixed values of the tangles.
Section ~\ref{concls} is devoted to the conclusions.

\section{Functional independence of $I_5$ and the tangles}\label{independentI5}

As in Refs.~\cite{Acin01,Gingrich02} we start from the canonical form
obtained by Ac\'{\i}n~\cite{Acin00}
\beq\label{Acinstate}
\lambda_1\ket{000}+\lambda_0\e^{i\varphi}\ket{100}+\lambda_2\ket{110}+
\lambda_3\ket{101}+\lambda_4\ket{111}
\eeq
with real positive $\lambda_i$ and $\varphi\in [0,\pi]$\footnote{We exchanged 
$\lambda_0$ with $\lambda_1$ and $\lambda_2$ with $\lambda_3$ 
respect to~\cite{Acin00}}.
This specific normal form is particularly convenient, since the threetangle $\tau_3$ assumes
the simple form 
\beq\label{tau3}
\tau_3=4 \lambda_1^2\lambda_4^2\ .
\eeq
It also permits an easy distinction of the two different SLOCC classes of 
entanglement.
The W-class corresponds to $\tau_3=0$, that is {\em iff} 
$\lambda_1=0$ or $\lambda_4=0$. In the former case, the state is a product.

The opposite extreme,
contained in the GHZ class, is when all concurrences vanish. Then, due
to the monogamy of entanglement for three qubits\cite{Coffman00},
all local density matrices are equivalent, since all local tangles 
$\tau_{1;i}:=4\det\rho_i$ coincide with $\tau_3$. We find
$4\det\rho_i=2\eps_{ijk}^2\lambda_i^2(\lambda_j^2+\lambda_k^2)+\tau_3+f_i$
with $\eps_{ijk}$ the Levi-Civit\'a tensor and 
$f_i=4\lambda_0\lambda_4(\lambda_0\lambda_4
   -2\cos\varphi\lambda_2\lambda_3)$ for $i\neq 1$ and $f_1=0$. 
These expressions are all equal to $\tau_3$
if $\lambda_0=\lambda_2=\lambda_3=0$. 

Taking a closer look at the independence of the six $SU(2)$ invariants reveals an
interesting connection between the functional independence of $I_5$ 
and the entanglement pattern of a state.
To this end we include the modulus $I_6:=\sum \lambda_i^2$ as the 6th invariant 
\beqa
\vec{Inv}&:=& (\tau_{1,1},\,\tau_{1,2},\,\tau_{1,3},\,\tau_3,\, I_5,\, I_6)\\
\vec{x}  &:=& (\lambda_0,\lambda_1,\lambda_2,\lambda_3,\lambda_4,\varphi)\ ,
\eeqa
and analyze the minors of $\nabla_{\vec{x}} \vec{Inv}$.
This yields that $I_5$ is functionally {\em dependent} of the remaining invariants
if either of the concurrences or the three-tangle vanishes. 
As two relevant examples we consider the case $C_{12}=0$ and $\tau_3=0$. 
For $C_{12}=0$ we find the relation
\beq\label{I5oftangles}
I_5=|\psi|^2\left(|\psi|^4-\frac{3}{4}\tau_{1,2} \right) \; ,
\eeq
which is maximized if $\tau_{1,2}=0$ $\Rightarrow$ $I_5=1$ 
and minimized for the GHZ state, i.e. if 
$\tau_{1,2}=1$ $\Rightarrow$ $I_5=\frac{1}{4}$.
For states from the W-class ($\tau_3=0$) we obtain
\beq\label{I5ofconcs}
I_5=|\psi|^6-\frac{3}{4} |\psi|^2(C_{12}^2+C_{13}^2+C_{23}^2)
    +\frac{3}{4}C_{12}C_{13}C_{23}\; ,
\eeq
which can be shown to be minimized if all concurrences are equal.
The maximum is again $I_5=1$ for full product states and the minimum
value $I_5=\frac{2}{9}$ is assumed for the W state 
$(\ket{100}+\ket{010}+\ket{001})/\sqrt{3}$; 
this can be demonstrated to be the absolute minimum.

Before going ahead to the analysis of $I_5$ we briefly mention a 
further invariant discussed in this context, namely the Grassl invariant. 
It discriminates
$\ket{\psi}$ from $\ket{\psi^*}$~\cite{Acin01} and is therefore necessary
for a complete characterization of the $SU(2)$ orbits for pure three 
qubit states~\cite{Gingrich02}.
It is a complex $SU(2)$ invariant which, evaluated on 
the Ac\'{\i}n normal form, gives
\beqa\label{ReGrassl}
\Re I_G &=& \tau_3\lambda_1^2\left [\cos(2\varphi)(\lambda_0\lambda_2\lambda_3)^2+
            \left[\cos\varphi \lambda_0\lambda_2\lambda_3\lambda_4+\frac{1}{4}\lambda_4^2(1-2(\lambda_0^2+\lambda_1^2))\right] (1-2(\lambda_0^2+\lambda_1^2))\right ]\\
\Im I_G &=& -\tau_3 \sin\varphi \lambda_0\lambda_2\lambda_3\left\{
            2\cos\varphi\lambda_0\lambda_2\lambda_3+ \lambda_4(1-2(\lambda_0^2+\lambda_1^2))
	    \right\}\label{ImGrassl}
\eeqa
It is clear that both real and imaginary part of $I_G$ are 
functionally dependent of the six other $SU(2)$ invariants including 
a discrete invariant emphasized on in Ref.~\cite{Gingrich02}. 
Interestingly $I_G$ vanishes if $\tau_3=0$ and is hence only
relevant for the GHZ class. Turning around this argument makes
immediately visible that for $\tau_3=0$, $I_5$ (which is a function of
the tangles, the modulus and the Grassl invariant) must be a function
of the concurrences and the modulus only.
%%%%%%%%%%%%%%%%%%%%%%%%%%%%%
\begin{figure}[ht]
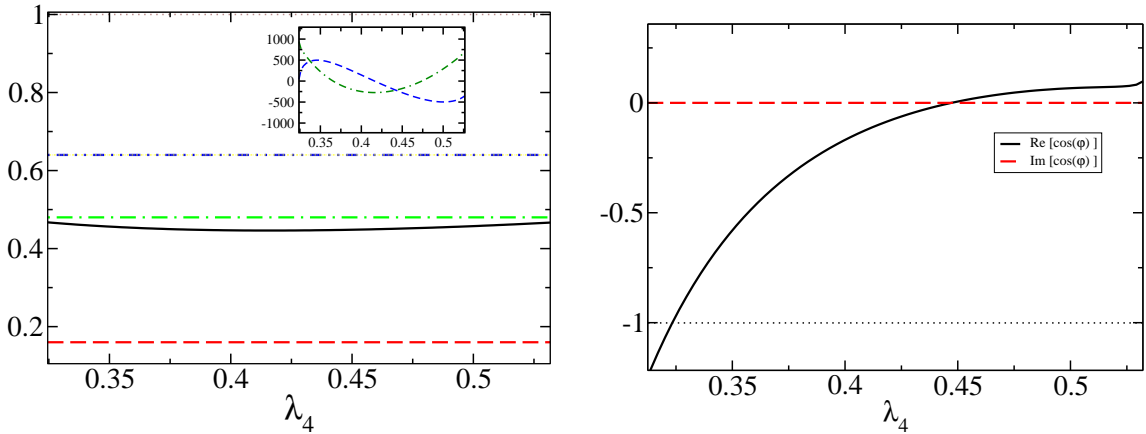

\begin{center}
\includegraphics[angle=0,width=0.48\textwidth]{I5.1111.pihalf.eps}
\hfill
\includegraphics[angle=0,width=0.48\textwidth]{Cosphi.1111.pihalf.eps}
\caption{The same as in figure \ref{Iso-pi} but for $\alpha=\pi/2$.
See legend of figure \ref{Iso-pi}. 
The inset displays the real and imaginary part of the
Grassl invariant $I_G$ (rescaled by $10^3$).
\label{Iso-pihalf}
}
\end{center}
\end{figure}
%%%%%%%%%%%%%%%%%%%

\section{One parameter family of Ac\'{\i}n states with fixed tangles}\label{main}

We now proceed with constructing the one-parameter families of states with all tangles held constant.
The variable $\lambda_4$ will be used as the free parameter in what follows. 
Starting from the relation Eq.~\eqref{tau3} for $\tau_3$, 
we successively obtain the following relations
\beqa\label{Result}
\lambda_1 &=& \bigfrac{\sqrt{\tau_3}}{2\lambda_4}\\
\lambda_2 &=& \bigfrac{\lambda_4}{\sqrt{\tau_3}}C_{12}\\
\lambda_3 &=& \bigfrac{\lambda_4}{\sqrt{\tau_3}}C_{13}\\
\lambda_0 &=& \sqrt{1-\bigfrac{\tau_3}{4\lambda_4^2}-\bigfrac{\lambda_4^2}{\tau_3}(C_{12}^2+C_{13}^2)-\lambda_4^2}
\eeqa
\beq\label{cosphi}
\cos(\varphi) = \bigfrac{4\lambda_4^2\tau_3-4\tau_3^2(\tau_3+C_{23}^2)
  + \lambda_4^4(C_{12}^2(4C_{13}^2-2\tau_3)-\tau_3(6C_{13}^2+5\tau_3))}
{4\lambda_4^2\tau_3 C_{12}C_{13}\sqrt{4\lambda^2_4-4\det\rho_1\frac{\lambda_4^4}{\tau_3}-\tau_3}}
\eeq
Among the concurrences and threetangle, only $C_{23}$ is $\varphi$ dependent.
If however, 
either $C_{12}=0$ or $C_{13}=0$, the phase $\varphi$ has no influence, 
neither on the tangles nor on $I_5$. 
In fact, the phase can be removed
by local phases  in both cases (e.g. $\e^{-i\varphi}$ in front of the basis element 
$\ket{1}$ of the first site and $\e^{i\varphi}$ in front of the basis 
element $\ket{1}$ of the second or third site, respectively).
\vspace*{8mm}
%%%%%%%%%%%%%%%%%%%%%%%%%%%%%
\begin{figure}[ht]
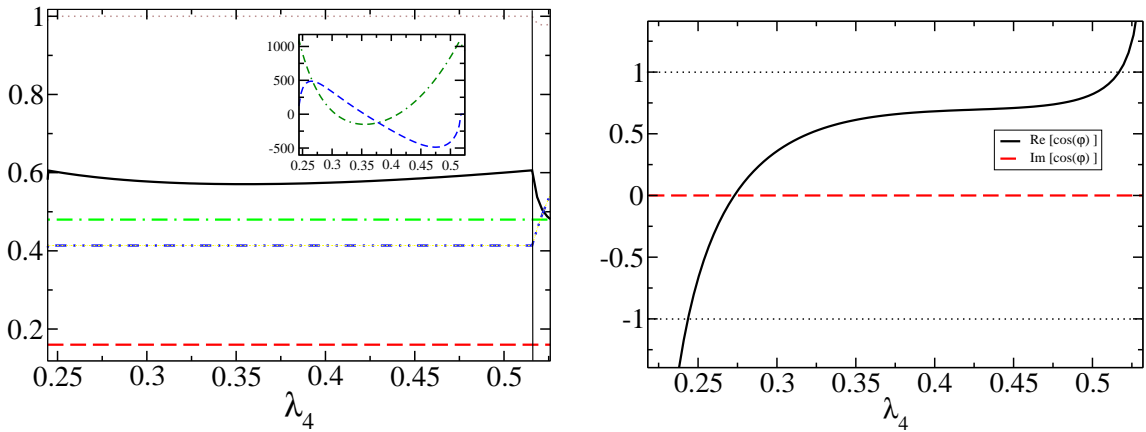

\begin{center}
\includegraphics[angle=0,width=0.48\textwidth]{I5.1111.pifourth.eps}
\hfill
\includegraphics[angle=0,width=0.48\textwidth]{Cosphi.1111.pifourth.eps}
\caption{The same as in figure \ref{Iso-pi} but for $\alpha=\pi/4$.
Here we show what typically happens beyond the interval of validity:
some local tangles deviate from that of the example state, whereas
$\tau_3$ is unchanged. Note that $\tau_3$ is the only invariant with respect
to the full $SL(2,\CC)^{\otimes 3}$. See legend of figure \ref{Iso-pi}.
The inset shows real and imaginary part of $10^3 \cdot I_G$.
\label{Iso-pifourth}
}
\end{center}
\end{figure}
%%%%%%%%%%%%%%%%%%%%%%%%%%%%%

Next we will discuss the interval accessible to $\lambda_4$.
Having a look at the square-root in the denominator of eq.(\ref{cosphi}),
we find that
\beq\label{l4range}
\lambda^2_4\in\bigfrac{\tau_3}{2\tau_{1,1}}\left\{
\begin{array}{rcl}
  \left[1-\sqrt{1-\tau_{1,1}},
               1+\sqrt{1-\tau_{1,1}}\right] & & \tau_{1,1}\in\left[0,\bigfrac{1}{2}\right]\\
 &\mbox{if}& \\
\left[1+\sqrt{1-\tau_{1,1}},
               1-\sqrt{1-\tau_{1,1}}\right] & & \tau_{1,1}\in\left[\bigfrac{1}{2},1\right]
\end{array}\right.
\eeq
Note that $\tau_{1,1}=C_{12}^2+C_{13}^2+\tau_3$.
Additional restrictions come from the zeros of $\lambda_0$ leading to
\beq\label{l4range2}
\lambda^2_4\in\bigfrac{\tau_3}{2(C_{12}^2+C_{13}^2)}\left\{
\begin{array}{rcl}
\left[1-\sqrt{1-C_{12}^2-C_{13}^2}, 1+\sqrt{1-C_{12}^2-C_{13}^2}
               \right] & &   C_{12}^2+C_{13}^2\leq\bigfrac{3}{4}\\
&\mbox{if}& \\
\left[1+\sqrt{1-C_{12}^2-C_{13}^2},1-\sqrt{1-C_{12}^2-C_{13}^2}
               \right] & &   C_{12}^2+C_{13}^2\geq\bigfrac{3}{4}
\end{array}\right.
\eeq
and besides the conditions $\lambda_i\leq 1$ for $i\in\{1,2,3\}$, leading to
\beq\label{l4range3}
\lambda_4\in\left[\sqrt{\tau_3}/2,\min\left\{\sqrt{\tau_3}/C_{12},\sqrt{\tau_3}/C_{13}\right\}\right]\; .
\eeq
In case of
three-tangle dominated entanglement, meaning here that $\tau_3$ is larger than
$C_{12}^2$ and $C_{13}^2$, the upper bound in ~\eqref{l4range3} is trivial.
It is worth
noticing that the predominance given to the first qubit is inherent
to the chosen normal form, which can as well be defined giving
the focus onto another qubit. The corresponding formulae would be
the same up to a permutation of the indices.

A further important restriction comes from  
$|\cos\varphi|\leq 1$; the latter bounds are obtained as the roots 
of a fourth order polynomial in $\lambda_4^2$ (see Eq.~\eqref{cosphi});
we will show $\cos\varphi(\lambda_4)$ in the figures.
\vspace*{8mm}
%%%%%%%%%%%%%%%%%%%%%%%%%%%%%
\begin{figure}[ht]
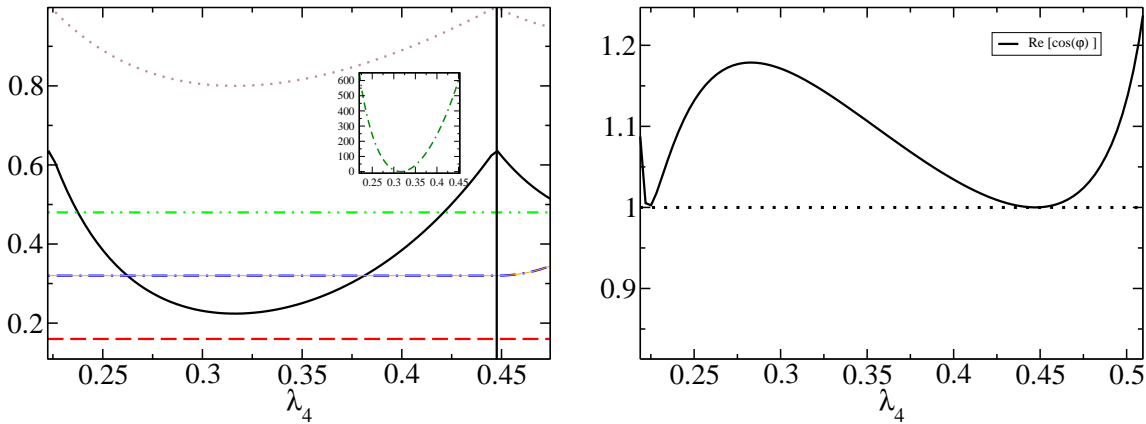

\begin{center}
\includegraphics[angle=0,width=0.48\textwidth]{I5.1111.0.eps}
\hfill
\includegraphics[angle=0,width=0.48\textwidth]{Cosphi.1111.0.eps}
\caption{The same as in figure \ref{Iso-pi} but for $\alpha=0$ and
consequently $C_{23}=0$. $I_5$ is hence no longer independent of the 
remaining $SU(2)$ invariants but given by Eq.~\eqref{I5oftangles}.
The ``interval'' of validity here consists of only two distinct points,
where $\cos \varphi=1$ and $|\psi|=1$. 
In between, the tangles are constant. 
Only $\tau_3$ is fixed all over the plot range, whereas the local
tangles start to vary beyond. $I_5$ varies all over the interval. 
See legend of figure \ref{Iso-pi}. 
The inset shows the Grassl invariant (rescaled by $10^3$), which is
real in this case.
\label{Iso-0}
}
\end{center}
\end{figure}
%%%%%%%%%%%%%%%%%%%%%%%%%%%%%

We illustrate our result using a reference state 
\beq\label{example}
\psi_\alpha=(\ket{000}+\e^{i\alpha}\ket{100}+
\ket{101}+\ket{110}+\ket{111})/\sqrt{5}
\eeq
for different values of the phase $\alpha$. It is clear that
all these states have the same threetangle but in general they differ
in their distribution of bipartite entanglement. 
The values for the tangles are $\tau_3=8/25$, $C_{12}=C_{13}=2/5$, and
$C_{23}=\sqrt{8}(1-\cos\alpha)/5$.
Then, the Kempe invariant $I_5$
is calculated, tuning through the one-parameter family of Ac\'{\i}n states
with the same entanglement pattern  as $\psi_\alpha$
(i.e. the same three concurrences $C_{12},C_{13},C_{23}$ and the same 
three-tangle $\tau_3$).
This one-parameter family of states is obtained from 
Eq.~(\ref{Acinstate}) subject to the replacements given in (\ref{Result})
- (\ref{cosphi}).
The result is plotted 
for $\alpha=\pi$ (fig.\ref{Iso-pi}), $\pi/2$ (fig.\ref{Iso-pihalf}), 
$\pi/4$ (fig.\ref{Iso-pifourth}), and for $\alpha=0$ (fig.\ref{Iso-0}).
In cases where the plot range exceeds the interval accessible to $\lambda_4$, its 
upper bound is indicated by a vertical line.
It is nicely seen that all measures for bipartite and tripartite
entanglement are constant within this interval. 
This implies that all states belong to the same non-zero
normal form under SLOCC local filtering operations\cite{VerstraeteDM03}.
As to be expected from the known functional independence of the 
$6$ local unitary invariants, $I_5$ varies continuously 
over the whole range - where also the normalization is preserved.
The interval admissible for $\lambda_4$ varies with the 
initial phase $\alpha$ (see figures).
Exceeding the upper bound of the accessible interval for $\alpha=\pi/4$ 
(fig.~\ref{Iso-pifourth}) 
it is seen that the threetangle $\tau_3$ remains fixed but both norm and some 
concurrences are no longer constant. This indicates that the states beyond
this limit are still $SL(2,\CC)^{\otimes 3}$ equivalent to the reference 
state but no longer norm preserving.
Within the admissible interval, the states are connected 
by norm-preserving $SL(2,\CC)^{\otimes 3}$ transformations 
(except for figure \ref{Iso-0}) but inequivalent with respect to local 
$SU(2)$ operations. 

\begin{figure}[ht]
\begin{center}
\includegraphics[angle=0,width=0.48\textwidth]{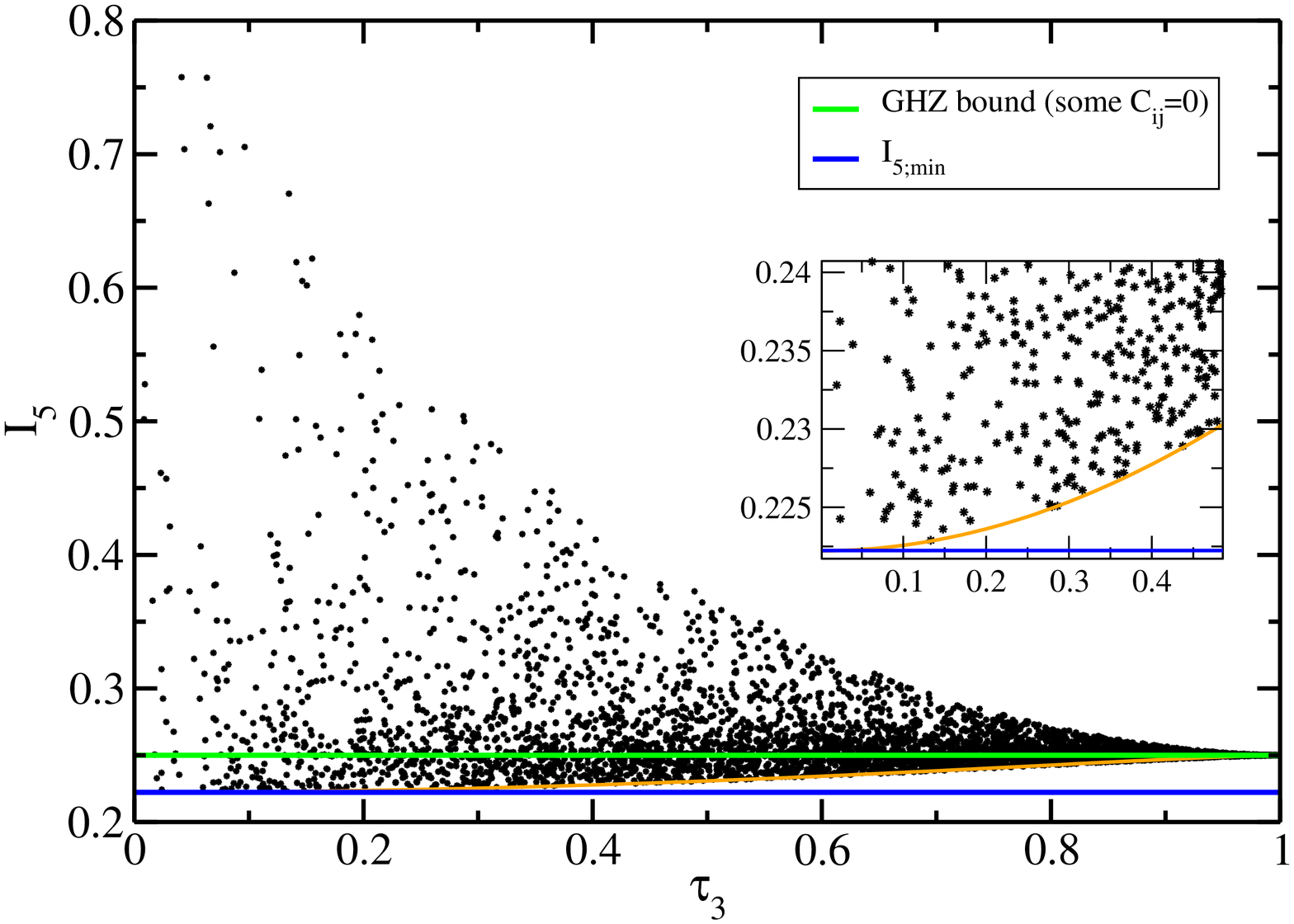}
\hfill
\includegraphics[angle=0,width=0.48\textwidth]{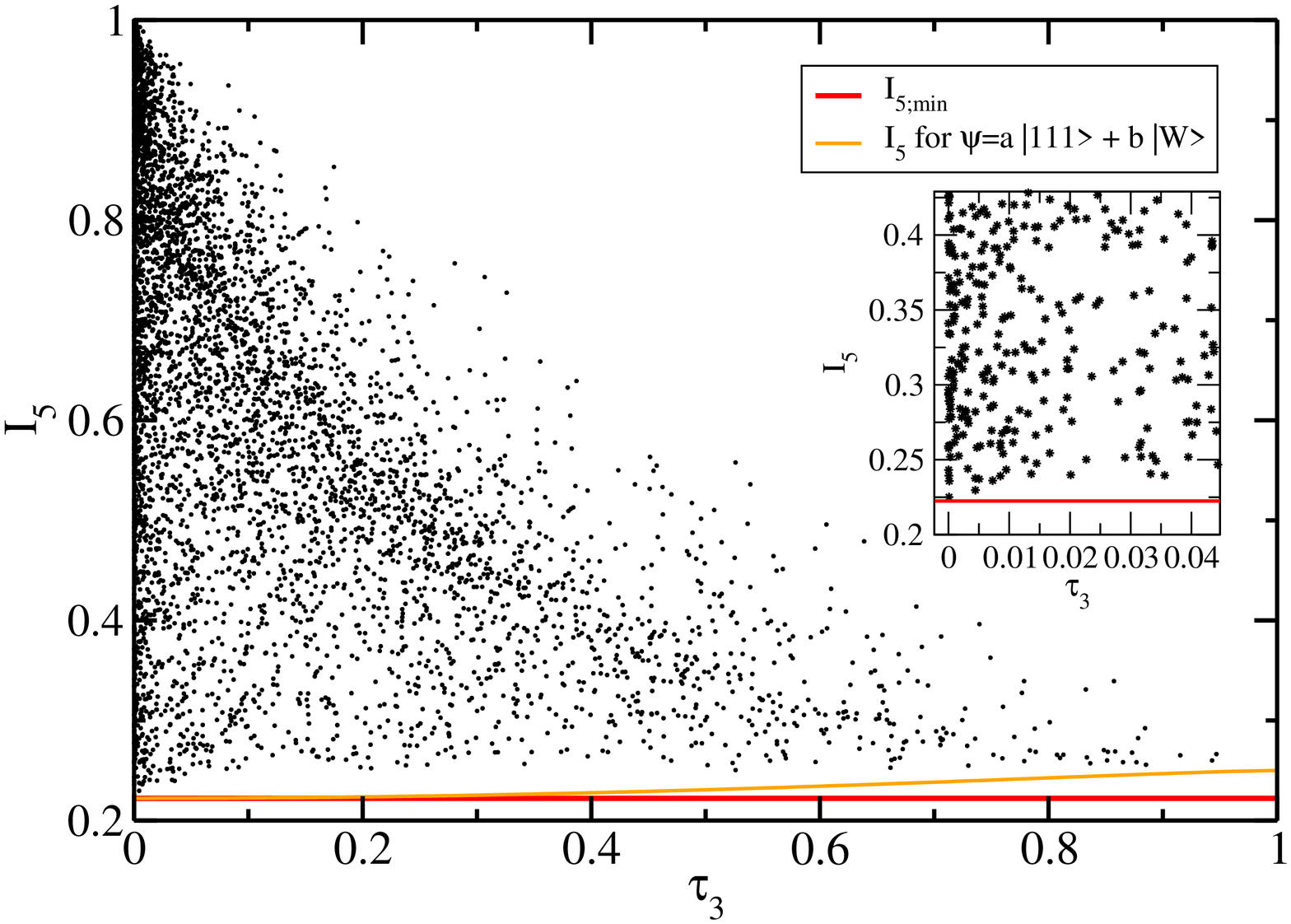}
\caption{{\em Left}: 
$I_5$ as a function of $\tau_3$ for a set of $5000$ random states 
in the GHZ-class (left panel) and for random states in the Ac\'{\i}n 
normal form (right panel).
Below the {\em GHZ bound} all concurrences are non-zero.
Obviously, due to the functional independence of $I_5$, it is not only that
$I_5$ varies when all tangles are kept constant, but also vice versa
can the tangles take wide ranges of values for fixed $I_5$.
The absolute minimum of $I_5$ as a function of $\tau_3$ is achieved for
states of the form $\ket{\psi}=a\ket{111}+\sqrt{1-a^2}\ket{W}$ (orange curve). 
\label{i5oftau3}
}
\end{center}
\end{figure}

Vice versa, it is clear that keeping $I_5$ fixed admits for a wide
range of the remaining tangles. This is seen in fig.~\ref{i5oftau3},
where we plot $I_5$ over $\tau_3$ for an ensemble of $5000$ 
random states out of a specific class. The left picture shows 
random states out of the GHZ class. The full red and green horizontal lines
indicate the absolute minimum $I_{5,\rm min}=2/9$
and what we called the GHZ bound for $I_5$.
The GHZ bound $1/4$ is the lower bound for states with at least one of 
the concurrences vanishing, and equation~\eqref{I5oftangles} applies. 
Below this bound, i.e. for $I_5<1/4$, the states are W-like in the sense that
all concurrences are positive. The right picture in fig.~\ref{i5oftau3}
shows an ensemble of $5000$ random Ac\'{\i}n states~\eqref{Acinstate}. It is seen
that for $\tau_3=0$ the Kempe invariant varies over its full range from
$2/9$ up to $1$.

%%%%%%%%%%%%%%%%%%%%%%%%%%%%%
\vspace*{5mm}
%%%%%%%%%%%%%%%%%%%%%%%%%%%%%
It is clear that local $SL(2,\CC)$ operations generate a drift in 
parameter space of the Ac\'{\i}n normal form. For norm-preserving transformations
and with all tangles larger or equal than some given $\epsilon>0$,
this leads to compact and connected 
$SL^{\otimes 3}$ orbit for each point in parameter space. 
Modulo the $SU(2)^{\otimes 3}$ invariance, 
each $SL(2,\CC)$ operation can be written in the form
\beq
\Matrix{cc}{s_1&r_1\\0&\frac{1}{s_1}}\otimes
\Matrix{cc}{s_2&r_2\\0&\frac{1}{s_2}}\otimes
\Matrix{cc}{s_3&r_3\\0&\frac{1}{s_3}}
\eeq
with real parameters $s_i,r_i$.
For norm preserving transformations, this leads to a five dimensional orbit
on the $5$-dimensional parameter space of normalized Ac\'{\i}n states.
Therefore, all states with equal $\tau_3$ and equal characteristic vector
$(\Theta(C_{12}),\Theta(C_{13}),\Theta(C_{23}))$ of non-zero 
concurrences~\footnote{$\Theta$ is the Heaviside step function}
are $SL^{\otimes 3}$ equivalent.
As an illustrative example consider the $SL(2)$ transformations
$\diag\{t,t^{-1}\}_1$ and  $\diag\{s,s^{-1}\}_2$, with $s,t\in\RR$
and the indices indicating the qubit number the matrix acts on.
Such diagonal transformations leave the Ac\'{\i}n state form-invariant.
If we want to keep the state normalized, this leads to a constraint
$s=s(t)$. Namely,
\beq
|s|^2=
\frac{1}{2}\frac{|t|^2\pm\sqrt{|t|^4(1-4\lambda_1^2(\lambda_2^2+\lambda_4^2))
    -4(\lambda_2^2+\lambda_4^2)(\lambda_0^2+\lambda_3^2)}}{\lambda_1^2 |t|^4+\lambda_0^2+\lambda_3^2}
\ ;\quad |t|\leq\sqrt[4]{\frac{4(\lambda_2^2+\lambda_4^2)(\lambda_0^2+\lambda_3^2)}{1-4\lambda_1^2(\lambda_2^2+\lambda_4^2)}}
\eeq
It is worth mentioning that even if $s$ is complex, a local relative
phase on the second qubit restores the original Ac\'{\i}n form.
Such transformations leave the threetangle and $C_{12}$ constant. 
When the initial state is taken as the reference state~\eqref{example}, 
even $C_{23}$ is constant, but this is a coincidence for that
particular state. It would be interesting 
to contruct explicitely those orbits with constant tangles, 
although their existence is clear from dimensional analysis together 
with the fact that $\tau_3$ is the only continuous polynomial 
$SL(2,\CC)^{\otimes 3}$ invariant. We leave this for future work. 

\section{Conclusions}\label{concls}
We have studied the Kempe-invariant $I_5$ for fixed entanglement pattern
in a tripartite qubit state. Its known functional independence in 
particular implies that $I_5$ will vary in general even when the 
entanglement pattern of the state, as given by the concurrences
and the threetangle, is kept fixed. This is indeed what is 
seen in the figures for some representative examples:
for fixed nonzero concurrences and threetangle, 
$I_5$ varies continuously over a finite interval.

As a further result we find that $I_5$ is functionally
independent only for globally distributed pairwise
entanglement, as is present for a $W$-state, in coexistence with 
the threetangle. So it looses its independent justification 
as soon as one of the tangles is zero. 
This already admits the conclusion that although $I_5$
is needed for a complete characterization of the local 
$SU(2)$ orbit of a state, 
it does not qualify for being an independent measure of entanglement.
In particular is $I_5$ not an additional measure of the distribution
of pairwise entanglement as surmised in Ref.~\cite{Sudbery01}.
Indeed, all entanglement measures for two qubits are equivalent to the 
concurrence, unless one wants to deviate from the mixed state extension
via the convex-roof. 
This uniqeness of the two-qubit entanglement measure is however also 
reflected in that all two-qubit states can be 
generated from a single Bell state~\cite{MREGS}. 
We can also exclude that $I_5$ be an additional measure for the 
global SLOCC class of entanglement.
The key observation to see this is that $I_5$ 
(besides the modulus of the state) is qualitatively different from
the remaining four $SU(2)$ invariants in that it has no $SL(2,\CC)$ invariance
on any of the three qubits. 
Therefore, local filtering operations~\cite{VerstraeteDM03} modify the 
value of $I_5$.
Under suitable (possibly infinitely many) local filtering operations
on three qubits the Kempe invariant, $I_5$, flows to its value for 
a normal form without concurrence.
In this limit it is functionally dependent of 
the modulus of the state and the threetangle.
In particular do states that differ only by their values of 
$I_5$ belong to the same SLOCC-entanglement class\cite{Duer00}
(having the same normal form). 

Furthermore, the $SL(2,\CC)$ action leads to compact orbits
acting continuously on the Ac\'{\i}n normal form, 
if only we bound all concurrences from below by some $\epsilon>0$. 
So in absence of some hypothetical discrete $SL(2,\CC)$ 
invariant that distinguishes different SLOCC inequivalent GHZ-classes
(see~\cite{Gingrich02} for $SU(2)$),
each two states with the same threetangle are $SL(2,\CC)^{\otimes 3}$ 
inter-convertible.
Assuming that $I_5$ incorporated such a hypothetical discrete $SL(2,\CC)$ 
invariant, 
it then should vary discontinuously with piecewise constant parts.
This is not what we observe. Hence, we can exclude this hypothesis.
This means that a continuous family of $SU(2)^{\otimes 3}$ inequivalent
states exist which are inter-convertible by norm preserving
$SL(2,\CC)^{\otimes 3}$ transformations.
Consequently, $I_5$ is not a measure for entanglement, since it
does not carry information about the entanglement structure of a 
three qubit quantum state (unless it is functionally dependent on the tangles).
It is worth adding here that we have statistical numerical evidence for 
$I_5$ being even an entanglement monotone: randomly chosen SLOCC operations 
containing up to $2$ Krauss operators on up to two qubits simultaneously
did not produce non-monotonic behavior. We therefore believe that
$I_5$ is even an entanglement monotone. 
This would mean that not even the monotone property~\cite{MONOTONES}
(which includes $SU(2)$ invariance) would be a 
conclusive
criterion for a quantity to being
useful for the classification and hence quantification of entanglement. 

In the light of the peculiarities encountered when dealing with convex-roof 
extended entanglement measures~\cite{LOSU,EOSU}, we also analyzed a 
possible connection to the entanglement of assistance
$E_a$ on two qubits which is given by the Uhlmann fidelity, the counterpart of 
the convex-roof extended concurrence. 
We find that $E_a$ can be expressed in terms of the tangles alone as
$E_{a,ij}=\sqrt{C_{ij}+\tau_3}$ without any connection to the Kempe invariant.

In order to clarify possible relevance of $I_5$ for quantum information 
processes not related to the entanglement of the state further analysis
is needed. The above line of arguments applies in the very same way to the 
Grassl invariant $I_G$.

Summarizing, a full classification of the local unitary orbits is
neither necessary nor sufficient for a classification of entanglement
for three qubits. This conclusion can be extended also to more qubits
and higher local Hilbert space dimensions, since the
classes of entanglement for three qubits appear as subclasses also there.
In contrast, those local unitary invariants that 
are also invariant under the local action of $SL(2,\CC)$
have proved to be necessary and sufficient for a full classification
for two and three qubit entanglement.
It has been demonstrated that $SL(2,\CC)$ invariants
give access to the classification of entanglement in multi-qubit
quantum states~\cite{OS04,OS05}. 
We have shown here that this requirement
can not be relaxed to $SU(2)$ invariance.

\acknowledgments
I would like to thank V. Giovannetti, 
R. Sch\"utzhold, J. Siewert, A. Uhlmann, and W. K. Wootters for 
fruitful discussions. I would like to thank R. Fazio for discussions and 
warm hospitality at the SNS Pisa in 2005, where the basis of 
this work has been set.

\end{document}